# Revisiting the prediction of solar activity based on the relationship between the solar maximum amplitude and max–max cycle length

V.M.S. Carrasco[1], J.M. Vaquero[2,3], and M.C. Gallego[1,3]

[1] Departamento de Física, Universidad de Extremadura, Badajoz, Spain (vmscarrasco@unex.es)

[2] Departamento de Física, Universidad de Extremadura, Mérida (Badajoz), Spain

[3] Instituto Universitario de Investigación del Agua, Cambio Climático y Sostenibilidad (IACYS), Universidad de Extremadura, Badajoz, Spain

**Abstract.** It is very important to forecast the future solar activity due to its effect on our planet and near space. Here, we employ the new version of the sunspot number index (version 2) to analyse the relationship between the solar maximum amplitude and max–max cycle length proposed by Du (2006). We show that the correlation between the parameters used by Du (2006) for the prediction of the sunspot number (amplitude of the cycle, $R_m$, and max-max cycle length for two solar cycles before, $P_{max-2}$) disappears when we use solar cycles prior to solar cycle 9. We conclude that the correlation between these parameters depends on the time interval selected. Thus, the proposal of Du (2006) should definitively not be considered for prediction purposes.

# 1. Introduction

The prediction of future solar activity levels is an important challenge owing to the impact they can have on Earth and our surrounding space. During the planning of Earth-orbit satellite missions, the solar activity indices are a tool employed to evaluate the future behaviour of solar activity (Mugellesi and Kerridge, 1991). Along with other parameters, these indices are used as input in an atmospheric model to predict the orbital decay. The greatest uncertainty for this model is found in the errors associated with the predicted values of the solar activity. We would emphasize that solar and geomagnetic activity can cause major problems in daily life due to our dependence on technological systems. Some effects of this phenomenon are saturation of transformers, disturbances in communication systems, corrosion in pipelines, etc. (Lanzerotti, 2001; Pulkkinen, 2007).

The attempts to predict the level of solar activity can be grouped in several ways (Petrovay, 2010). Pesnell (2012) presented a summary of 75 predictions of the maximum amplitude for the current solar cycle 24, classified into different categories – climatology, dynamo models, spectral, etc. Since the sunspot number series is the longest observational series available (Clette et al., 2014), it is the index most used to predict solar activity. However, several authors (Hathaway et al., 1999; Petrovay, 2010; Pesnell, 2012) have noted that prediction methods based on cycle characteristics from sunspot numbers are less reliable than methods based on geomagnetic precursors or polar fields (Svalgaard et al., 2005).

Two notable parameters related to the solar cycle used in prediction tasks are the solar maximum amplitude ($R_m$) and the cycle length ($P$). Several relationships have been found involving these parameters. Du (2006) proposed a prediction method based on the relationship between $R_m$ and $P_{max}$ (cycle length defined from maximum to maximum) at lag -2 after analyzing the correlations of these parameters at different lags for the old version of the sunspot number (version 1) considering data from solar cycle 9 onwards. Carrasco et al. (2012) demonstrated that previous solar cycles can be considered for the analysis because they have good temporal coverage, and showed that the correlation used by Du (2006) between $R_m$ and $P_{max}$ at lag –2 from solar cycle 6 onwards disappears, and therefore the proposal of Du (2006) must not be used for prediction purposes. Moreover, other authors (Hathaway et al., 1994; Solanki et al., 2002) found that the strongest correlation between the solar maximum amplitude and the cycle length defined from minimum to minimum is at lag –1.

There exist other well-known relationships between parameters of the solar cycle used for prediction tasks, for example, amplitude – rise time. This relationship is known as the "Waldmeier Effect" (Waldmeier, 1935), and it states that solar cycles with greater maximum amplitudes have shorter rise times. This effect is significant in the sunspot number series, but several authors have shown that the correlation between these parameters is weaker in the sunspot area series (Kane, 2008; Karak and Choudhuri, 2011; Carrasco et al., 2016).

Since certain problems have been detected in the sunspot number series, mainly in the historical part (Vaquero, 2007), they are being revised (Clette et al., 2014). As a consequence, new versions of the sunspot number have recently been published (Clette et al., 2015, Usoskin et al., 2016, Svalgaard and Schatten, 2016; Lockwood et al., 2016), and a new revised collection of sunspot group numbers is available (Vaquero et al., 2016). The objective of the present work is to revise the works of Du (2006) and Carrasco et al. (2012) using the new version of the sunspot number (http://www.sidc.be/silso/, version 2) to provide a definitive analysis of Du's prediction method. In Section 2, we present the methods employed in this revision (weighted average epochs of maxima, and Gaussian filter), and we analyse the relationship between $R_m$ and $P_{max}$ for all data of the new version of the sunspot number. The analysis and results are shown in Section 3 and, finally, Section 4 is devoted to the conclusions.

## 2. Methods

In order to analyse the relationship between the solar maximum amplitude and max–max cycle length, previous work used the old versions of the sunspot number index. While Du (2006) employed the international sunspot number (Clette et al., 2014), Carrasco et al. (2012) used both the international sunspot number and the group sunspot number (Hoyt and Schatten, 1998). For the present work, we used the new version of the sunspot number published recently (Clette et al., 2015) and available on the Web (www.sidc.be/silso/datafiles).

We performed the analysis in two ways. On the one hand, we followed the method of Du (2006) to calculate the weighted average epochs of maxima. For this purpose, we employed the 13-month smoothed values corresponding to the new version of the sunspot number. To obtain the epoch of the maximum of a given solar cycle, we selected all values lying within the range $R_m - d$ to $R_m$, with $R_m$ being the maximum value for the cycle and $d = 0.1 \cdot (R_m - R_0)$, where $R_0$ is the minimum value of the cycle. The maxima are defined by:

$$E_m = \frac{1}{\sum_{i=1}^{n} \omega_i} \sum_{i=1}^{n} E_i \omega_i$$

where $E_i$ are the dates and $\omega_i$ ($\omega_i = 1/(R_m-R_i)$) the weights of the range selected previously. Here, $R_i$ is the value of the sunspot number for each point of that interval, and, in the case of $R_i = R_m$, $\omega_i$ is taken as $3\omega'$, with $\omega'$ being the maximum weight for $R_i \neq R_m$ (Du et al., 2006a, 2006b). Finally, the max–max cycle length is calculated by $P_{max}(i) = E_m(i) - E_m(i-1)$, where $E_m(i)$ and $E_m(i-1)$ are the weighted average epochs of the maxima for cycles $i$ and $i-1$, respectively. Table 1 lists the parameters used in the analysis of the relationship between $R_m$ and $P_{max}$ following this method.

Table 1. Parameters used in this analysis for the 13-month smoothed sunspot number.

| CYCLE | $E_m$ | $R_m$ | $P_{max}$ | CYCLE | $E_m$ | $R_m$ | $P_{max}$ |
|---|---|---|---|---|---|---|---|
| 1 | 1761 Jun | 144.1 | – | 13 | 1893 Nov | 142.9 | 119 |
| 2 | 1769 Oct | 190.9 | 100 | 14 | 1906 Mar | 106.4 | 148 |
| 3 | 1778 Jun | 260.9 | 104 | 15 | 1917 Aug | 175.7 | 137 |
| 4 | 1788 Feb | 235.3 | 116 | 16 | 1928 Mar | 128.9 | 127 |
| 5 | 1804 Oct | 80.3 | 200 | 17 | 1937 May | 198.3 | 110 |
| 6 | 1816 May | 81.2 | 139 | 18 | 1947 Jul | 214.1 | 122 |
| 7 | 1829 Dec | 118.6 | 163 | 19 | 1958 Feb | 284.6 | 127 |
| 8 | 1837 Mar | 244.9 | 87 | 20 | 1969 Jan | 155.7 | 131 |
| 9 | 1848 Jun | 204.4 | 135 | 21 | 1980 Jan | 232 | 132 |
| 10 | 1860 Feb | 186.2 | 140 | 22 | 1989 Nov | 212.5 | 118 |
| 11 | 1870 Aug | 234 | 126 | 23 | 2001 Sep | 177.1 | 142 |

| 12 | 1883 Dec | 124.4 | 160 | 24 | 2014 Apr | 116.4 | 151 |

On the other hand, we applied a Gaussian filter to the sunspot number index in order to establish the maxima and the cycle lengths for this series. Hathaway (2015) showed that the 13-month running mean does not work well for high-frequency variations. In contrast, the Gaussian filters are preferable because they remove these variations. A suitable Gaussian filter is given by:

$$W(t) = e^{-t^2/2a^2} - e^{-2}(3 - t^2/2a^2)$$

with $-2a + 1 \leq t \leq +2a - 1$, where $t$ is the time from the centre of the filter and $2a$ is the Full Width at Half Maximum (FWHM) of the filter. The significant variations in solar activity on time scales of one to three years are filtered by a 24-month Gaussian filter. Table 2 shows the parameter used for the analysis applying the Gaussian filter. Note that: (i) it is not possible to calculate the cycle length for solar cycle 1 due to there being no previous data, and (ii) for the Gaussian filter, there is not enough data to calculate the parameters corresponding to solar cycle 24.

Table 2. Parameters used in this analysis for applying a 24-month Gaussian filter to the sunspot number.

| CYCLE | $E_m$ | $R_m$ | $P_{max}$ | CYCLE | $E_m$ | $R_m$ | $P_{max}$ |
|---|---|---|---|---|---|---|---|
| 1 | 1761 May | 121.4 | – | 13 | 1893 Sep | 149.1 | 118 |
| 2 | 1770 Jan | 167.6 | 104 | 14 | 1906 May | 106.6 | 152 |
| 3 | 1778 Sep | 229.1 | 104 | 15 | 1917 Dec | 186.8 | 139 |

| 4 | 1788 Mar | 217.7 | 114 | 16 | 1927 Dec | 136.7 | 120 |
| --- | --- | --- | --- | --- | --- | --- | --- |
| 5 | 1804 Jul | 76.1 | 196 | 17 | 1937 Dec | 199.7 | 120 |
| 6 | 1816 Aug | 72.9 | 145 | 18 | 1948 Jan | 228.3 | 121 |
| 7 | 1829 Oct | 111.8 | 158 | 19 | 1958 Feb | 295.0 | 121 |
| 8 | 1837 Mar | 254.7 | 89 | 20 | 1969 Mar | 157.9 | 133 |
| 9 | 1848 Aug | 223.6 | 137 | 21 | 1980 May | 236.6 | 134 |
| 10 | 1860 Mar | 187.1 | 139 | 22 | 1990 May | 217.9 | 120 |
| 11 | 1870 Nov | 241.5 | 128 | 23 | 2001 May | 185.1 | 132 |
| 12 | 1883 Nov | 130.1 | 156 | 24 | – | – | – |

The analysis of Du (2006) included data from solar cycle 9 to 23, arguing that these cycles are the most reliable for the sunspot number data. However, Carrasco et al. (2012) demonstrated that solar cycles 6, 7, and 8 have good temporal coverage, and therefore should be considered for the analysis. In this present study, we analysed the relationship between the solar maximum amplitude and max–max cycle length given for the two methods presented above from: (i) solar cycle 2, (ii) solar cycle 6, and (iii) solar cycle 9, onwards. Thus, we reproduce the previous studies of Du (2006) and Carrasco et al. (2012) with the new version of the sunspot number. Furthermore, we examine the relationship between the two parameters studied from the beginning of the series to analyse the relationship between $R_m$ and $P_{max}$ for the new version of the sunspot number.

## 3. Analysis and Results

First of all, we calculated the different correlation coefficients using different lags between the parameters $R_m$ and $P_{max}$ (from lag 0 to –6). Figure 1 represents these correlation coefficients according to (top panel) the weighted average epochs of the maxima and (bottom panel) using the Gaussian filter. Blue, red, and purple colours indicate the results obtained from solar cycles 2, 6, and 9 onwards, respectively.

Taking into account the weighted average epochs of the maxima and all solar cycles, the highest values for the correlation corresponded to lag 0 ($r = 0.692$, $p$-value $< 0.001$), while at lag –2 the correlation was $r = 0.186$, $p$-value $= 0.420$. Considering data from solar cycle 6 onwards, we also obtained the stronger correlation at lag 0 ($r = 0.607$, $p$-value $= 0.006$), and, for lag –2, $r = 0.233$, $p$-value $= 0.368$. Using only data from solar cycle 9 onwards, the strongest correlation was again at lag 0 ($r = 0.525$, $p$-value $= 0.037$). However, in this case, the correlation at lag –2 was similar to that at lag 0 ($r = 0.502$, $p$-value $= 0.067$).

With the Gaussian filter, considering all solar cycles, we obtained the strongest correlation for the lag 0 ($r = 0.667$, $p$-value $< 0.001$), while at lag –2 the correlation coefficient was $r = 0.130$, $p$-value $= 0.585$. From solar cycle 6 onwards, the highest value of the correlation was at lag 0 ($r = 0.624$, $p$-value $= 0.006$), and for lag –2, $r = 0.223$, $p$-value $= 0.410$. Using data from solar cycle 9 onwards, the correlation coefficient at lag 0 was $r = 0.468$, $p$-value $= 0.079$. However, the highest value was found at lag –2 ($r = 0.605$, $p$-value $= 0.028$).

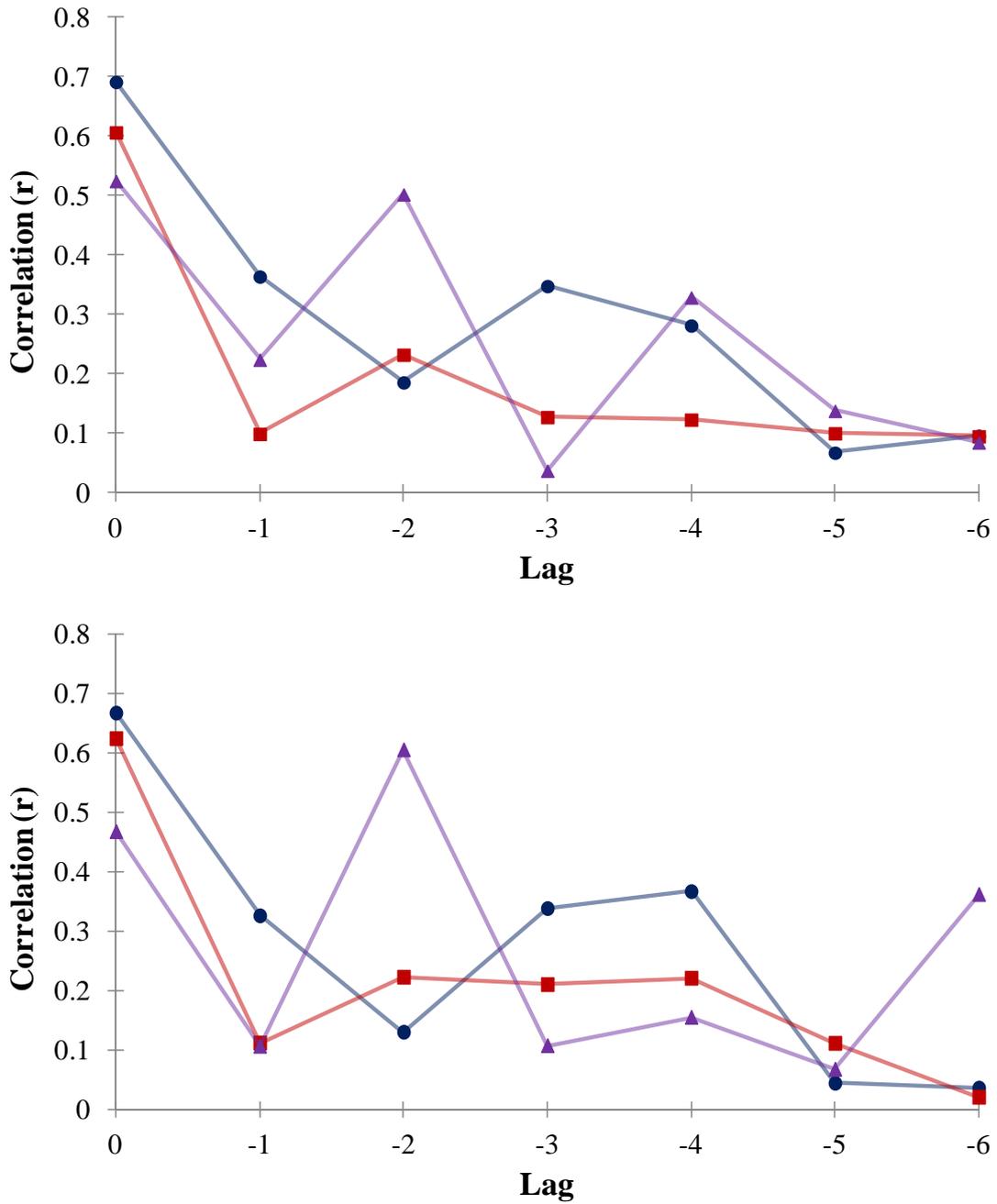

Figure 1. Correlation coefficients corresponding to the relationship between $R_m$ and $P_{max}$ from lag 0 to –6 according to the weighted average epochs of maxima (top panel) and the Gaussian filter (bottom panel). Blue circles, red squares, and purple triangles represent the analyses from solar cycles 2, 6, and 9 onwards, respectively. Data were extracted from the new version of the sunspot number (www.sidc.be/silso/datafiles).

This analysis shows that the strongest correlations are achieved between the cycle maximum amplitude and the length of the same cycle (lag 0), taking into account all the sunspot number data. These results are similar to those obtained by Carrasco et al. (2012). However, they differ from the results of Du (2006) who found the strongest correlation for the relationship between the parameters $R_m$ and $P_{max-2}$. In our analysis, the correlation between $R_m$ and $P_{max-2}$ has the highest values with the data from solar cycle 9 onwards using the Gaussian filter ($r = 0.605$, $p$-value $= 0.028$). Nevertheless, Carrasco et al. (2012) demonstrated that several solar cycles prior to solar cycle 9 have good temporal coverage, and should be considered for the analysis. Moreover, this result does not agree with other works using the old version of the sunspot number (Hathaway et al., 1994; Solanki et al., 2002) which obtain the strongest correlation between the maximum amplitude and the length of the preceding cycle, although, unlike the present work, they defined the cycle length from minimum to minimum.

It can be seen that the correlation according to the method of Du (2006) gets its strength from two solar cycles, cycles 12 and 17. We therefore calculated the different correlation coefficients following Du's method, but removing the parameters ($R_m$ and $P_{max}$) belonging to a single solar cycle. This gave us sets of 14 (for the weighted average epochs of maxima) and 13 (for the Gaussian filter) correlation coefficients (Figure 2). One observes in the figure that solar cycle 17 for the weighted average epochs of maxima and solar cycle 12 for Gaussian filter lead to values that lie outside the limits of mean-value-plus-two-standard-deviations represented by the dashed lines. Moreover, solar cycles 12 and 22 also lie outside the limit of mean-value-plus-one-standard-deviation for the weighted average epochs of maxima, as do solar cycles 9 and 18 for the Gaussian filter.

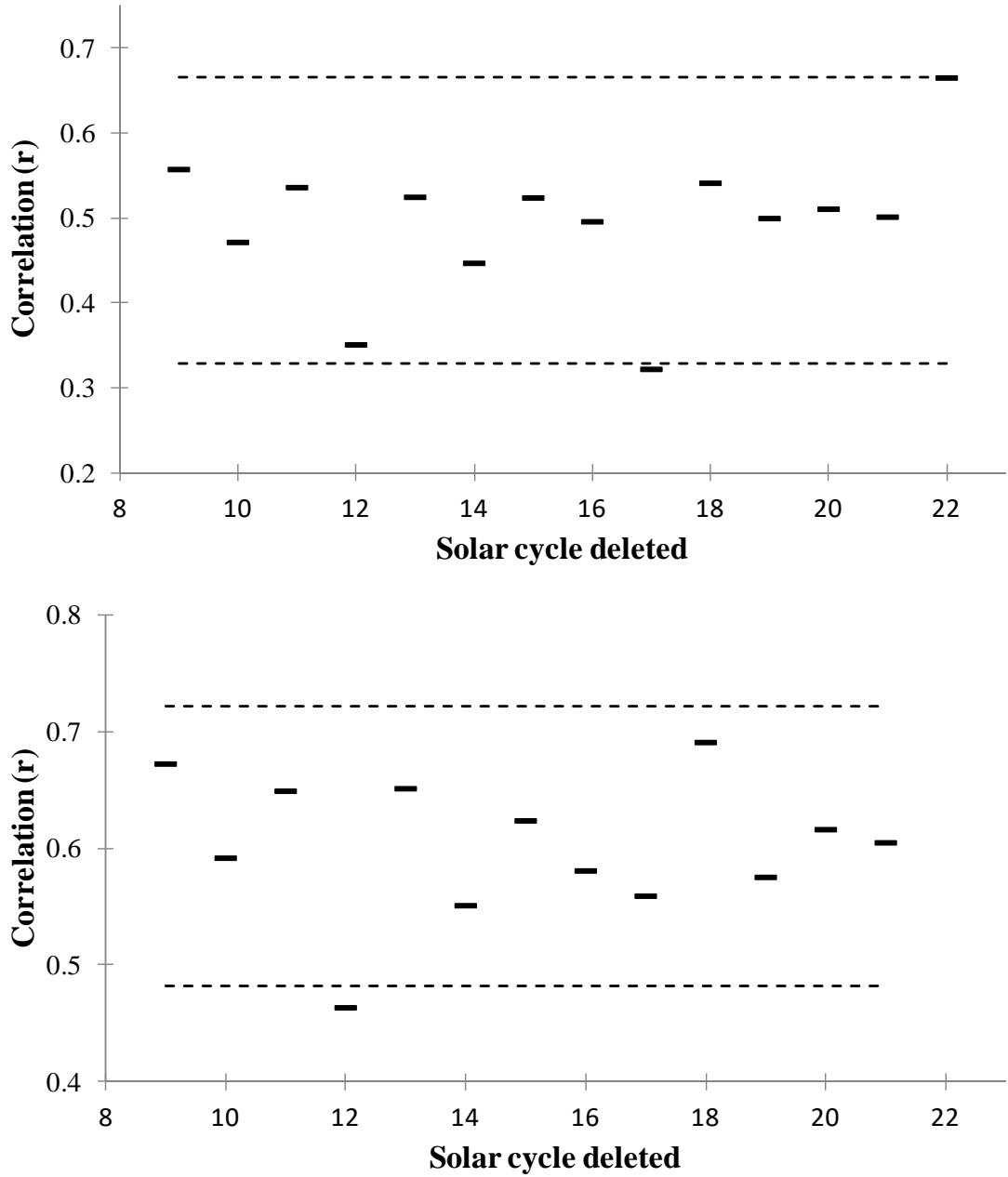

Figure 2. Correlation coefficients calculated when $R_m$ and $P_{max}$ of a single solar cycle are removed following the analysis of Du (2006). Top panel (bottom panel) shows the calculation according to the weighted average epochs of maxima (Gaussian filter). The dashed lines represent the mean value plus/minus two standard deviations.

## 4. Conclusions

We have re-analysed the relationship between the amplitude ($R_m$) and max-max length ($P_{max}$) of the solar cycle previously studied by Du (2006) and Carrasco et al. (2012). For this purpose, we used the data from the new version of the sunspot number (version 2). To define the maxima of the solar cycles, two different methods were used: (i) the weighted average epochs of maxima following Du (2006), and (ii) a 24-month Gaussian filter.

In order to reproduce the aforementioned two works, we made analyses from solar cycles 6 and 9 onwards. Moreover, to investigate the behaviour of the relationship between $R_m$ and $P_{max}$ in the new sunspot number series, we performed another test from the beginning of this series (from solar cycle 2), calculating the different correlation coefficients from lag 0 to –6.

Generally, we found the strongest correlations to be between the amplitude of a solar cycle and its own length with both the weighted average epochs and the Gaussian filter from solar cycles 9, 6, and 2 onwards. However, the strongest correlation for the analysis with the Gaussian filter from solar cycle 9 onwards was between $R_m$ and $P_{max-2}$, as Du (2006) had obtained. Specifically, the strongest correlations were obtained taking into account all the data of the sunspot number series with both the weighted average epochs ($r = 0.692$, $p$-value $< 0.001$) and the Gaussian filter ($r = 0.667$, $p$-value $< 0.001$). These values improve the correlation coefficients calculated for data from solar cycle 6 onwards (weighted average epochs: $r = 0.607$, $p$-value $= 0.006$; Gaussian filter: $r = 0.624$, $p$-value $= 0.006$).

The present results calculated from the new version of the sunspot number agree with those obtained in Carrasco et al. (2012). Also, the correlation between $R_m$ and $P_{max-2}$ practically disappeared when we carried out the analysis from solar cycle 6 and the beginning of the sunspot number series onwards. The relationship between these parameters depends on the temporal interval, and seems to be more complex than a simple lag. Therefore, after analysing the proposal of Du (2006) with the new version of the sunspot number, we can definitively conclude that this method must not be employed for prediction tasks. Note that the prediction of Du (2006) for solar cycle 24 was 150.3 ± 22.4, while the maximum official value according to the smoothed sunspot number series (version 1) for that cycle is 81.9.

In view of the results, the solar dynamo seems to have a certain memory with respect to prior solar cycles, not greater than one solar cycle, since the correlations between the two parameters analysed here generally decrease as the lags are made greater.

## Acknowledgements

Support from the Junta de Extremadura (Research Group Grant GR15137) and from the Ministerio de Economía y Competitividad of the Spanish Government (AYA2014-57556-P) is gratefully acknowledged.